\documentclass[5p]{elsarticle}
\usepackage{amsmath,amsfonts,amssymb}
\usepackage{bm}
\usepackage{algorithm}
\usepackage{algorithmic}
\usepackage{url}
\usepackage[colorlinks=true,linkcolor=blue,urlcolor=blue]{hyperref}
\usepackage{xcolor}
\usepackage{MnSymbol}
\colorlet{markercolor}{purple!50!black}
\def\B#1{\bm{#1}}

\begin{document}

\title{Markov models for fMRI correlation structure: is brain functional
connectivity small world, or decomposable into networks?}

\author[parietal,unicog,cea]{G.~Varoquaux\corref{corresponding}}
\ead{gael.varoquaux@inria.fr}
\author[parietal,cea]{A.~Gramfort}
\ead{alexandre.gramfort@inria.fr}
\author[cea]{J.B.~Poline}
\ead{jbpoline@cea.fr}
\author[parietal,cea]{B.~Thirion}
\ead{bertrand.thirion@inria.fr}

\cortext[corresponding]{Corresponding author}

\address[parietal]{Parietal project-team, INRIA Saclay-\^ile de France}
\address[unicog]{INSERM, U992}
\address[cea]{CEA/Neurospin b\^at 145, 91191 Gif-Sur-Yvette}

\begin{abstract}
    \noindent 
    Correlations in the signal observed via functional Magnetic Resonance Imaging (fMRI), are
    expected to reveal the interactions in the underlying neural
    populations through hemodynamic response.
    In particular, they highlight distributed set of mutually correlated regions
    that correspond to brain networks related to different cognitive functions.
    %
    %
    Yet graph-theoretical studies of neural connections give a 
    different picture: that of a highly integrated system with small-world
    properties: local clustering but with short pathways across the
    complete structure.
    %
    %
    We examine the conditional independence properties of the fMRI
    signal, \emph{i.e.} its \emph{Markov structure}, to find realistic
    assumptions on the connectivity structure that are required to explain 
    the observed functional connectivity.
    In particular we seek a decomposition of the Markov structure into 
    segregated functional networks using \emph{decomposable graphs}:
    a set 
    of strongly-connected and partially overlapping cliques. 
    We introduce a new method to efficiently 
    extract such cliques on a large, strongly-connected graph.
    We compare methods learning different graph structures from functional 
    connectivity by testing the goodness of fit of the model they
    learn on new data.
    We find that summarizing the structure as strongly-connected networks 
    can give a good description only for very large and overlapping
    networks. These results highlight that Markov models are good tools
    to identify the structure of brain connectivity from fMRI signals,
    but for this purpose they must reflect the small-world properties of
    the underlying neural systems.
\end{abstract}


\maketitle

\sloppy 

\providecommand{\OO}[1]{\mathcal{O}\bigl(#1\bigr)}
\providecommand{\B}[1]{\mathbf{#1}}

\section{Introduction}
\label{submission}

The study of distant correlations in functional Magnetic Resonance
Imaging (fMRI) signals, revealing brain \emph{functional connectivity},
has opened the door to fundamental insights on the functional
architecture of the brain \cite{fox2007,Bullmore2009}.
Among other features, the concept of cognitive network has emerged as one
of the leading views in current cognitive neuroscientific studies:
regions that activate together form an integrated network corresponding
to a cognitive function \cite{fox2007}. 
In other words, these networks can in general be identified to
sets of active regions observed in traditional brain mapping
experiments, thus in line with a segregative view of brain functional
organization.
On the other hand, graph-theoretical analysis of brain connectivity has shown that it
displays small-world properties: any two points of the brain can be 
connected through few intermediate steps,
despite the fact that most nodes maintain only a few direct connections
\cite{Bullmore2009}. 
There is an apparent contradiction between those two points of views: can
 the brain be separated in functionally-specialized networks, or are these
simply a convenient but misleading representation to interpret the data?


Establishing a link between the functional connectivity observed in
fMRI and the underlying neural architecture can be a challenging
task. Indeed, the fMRI signal is very noisy and reflects many non
neural effects, such as cardiac and respiratory rhythms or head and
body motion. 
\noindent
In addition, statistical estimation of brain interactions from fMRI
data is made difficult by the small number of observations available
per experimental run, and by the lack of salient structure in this
data.
\noindent
Finally, it has become a common practice to study these interactions
during \textit{resting-state}, i.e. in the absence of any stimulation,
the subject resting eyes-closed in the scanner; in that case functional
connectivity is loosely identified with the correlation structure of
ongoing activity across regions.

In this paper, we use advanced statistical techniques to investigate
the structure of brain connections that is captured by resting-state
fMRI recordings. For this purpose, we infer the Markov network
structure of the signals --the underlying independence graph-- using 
a statistical framework well suited to
modeling correlations: Gaussian graphical models. We compare various
estimation methods adapted to different graph
topologies. In particular, we introduce decomposable Markov models
to describe functional networks: sets of distant regions working together. 
Importantly these tools enable to
study the extent --i.e. the clique size-- of brain functional
networks that is required to give a good description of the signal.
Our work introduces a probabilistic description of brain activity that
encompasses features of distributed cognitive network and small-world systems.

The layout of the article is the following. First we review different
pictures of brain connectivity emerging from a vast body of anatomical
and structural studies. In section \ref{sec:statistical_tools}, we
introduce the statistical tools that we use to estimate functional
connectivity graphs from the fMRI signal. In section
\ref{sec:algorithm}, a method for learning decomposable models of
brain activity is presented. Finally, in section
\ref{sec:experiments}, we estimate graphical models of brain activity
with various methods and quantify their ability to fit unseen data as
a function of the imposed graph topology: a strong segregation into
well-separated distributed networks, or a highly-connected system.

\section{Brain structure: from small-world connectivity to functional
networks}

\paragraph{Integration and segregation in neural systems}
Beyond the historical opposition of \emph{localizationist} and
\emph{holist} views \cite{finger2001}, brain organization is generally
considered as reflecting the fundamental principles of segregation into
functionally-specialized sub-systems, and integration across these
systems to sustain high-level cognitive functions \cite{tononi1998}.
Originally highlighted in theoretical descriptions of neural systems
\cite{tononi1994}, this general organization shapes the properties of
brain-wide connectivity: local clustering and
global connectivity \cite{sporns2004,Bullmore2009} that reflects
functional integration \cite{varoquaux2010c}.
In high-level cognition, it can been seen as the duality of local
circuits and global networks \cite{dehaene1998}, for instance in
conscious information integration \cite{bekinschtein2009}.
Finally, there is growing evidence from functional imaging, that this
segregation increases during brain development \cite{fair2007,fair2009}.

\paragraph{Small word graphs}
A quantitative description of these structural properties can be given
using graph theory. In mathematics and computer science, a \emph{graph}
is the abstract object defined by a set of nodes and their connections.
In social sciences or neuroscience, the word ``network'' is often
employed instead of ``graph'', but in this paper, we will reserve this
term to denote cognitive or functional networks.
\citet{watts1998} popularized the concept of \emph{small-world}
properties for graphs: graphs with a small fraction of nodes connected,
but with a connection topology such that only a short chain is required
to connect any pair of nodes. These efficient transport properties are
achieved via a local clustering into \emph{communities}, and a few edges
interconnecting the communities across the graph.
A graph can be characterized by a variety of statistical properties of
its connections \cite{strogatz2001,amaral2000,Bullmore2009}. Most often,
small-world graphs are studied via the relative importance of a local
graph metric, such as their clustering coefficient, and a global metric,
\emph{e.g.} their average path length
\cite{watts1998,strogatz2001,humphries2008}. The fundamental
characteristic of a small-world graph remains that it displays high
sparsity and yet good transport across the graph. \citet{watts1998}
stress that, for sparse graphs, small-worldness is determined by its
global properties: \emph{``at the local level (as reflected by [the
clustering coefficient]), the transition to a small world is almost
undetectable''} while \emph{``in sparse networks with many vertices [...]
even a tiny fraction of short cuts would suffice''}.

\paragraph{Graphs of brain connectivity} The physical connections
between gray-matter areas can be measured in-vivo via tractography of
diffusion MRI \cite{conturo1999,lebihan2003}. They establish a picture of
the human brain architecture at an anatomical-connectivity level 
\cite{sporns2005,hagmann2008}. In non-human animals,
systematic post-mortem studies have enabled detailed and complete mapping
of connections, \emph{e.g.} for the macaque brain
\cite{felleman1991,stephan2001}.
This anatomical connectivity forms the substrate of information
processing in the brain and reflects the functional segregation
properties of neural systems \cite{sporns2000}. From the point of view of 
graph theory, it has been shown to display small-world properties
\cite{sporns2004,Bullmore2009}. In fact, \citet{watts1998} used the graph
of neural connections of \emph{C.~elegans} in their seminal paper
introducing small-world graphs.
It is also possible to build graphs of functional connectivity 
by thresholding correlation matrices of brain activity, for instance from MEG
\cite{stam2004} or fMRI data
\cite{salvador2005,achard2006,bassett2006,eguiluz2005}. As their
anatomical counterparts, these have been shown to display small-world
properties: local clustering and short average path length.

\begin{figure}
\centerline{%
    \includegraphics[width=0.7\linewidth]{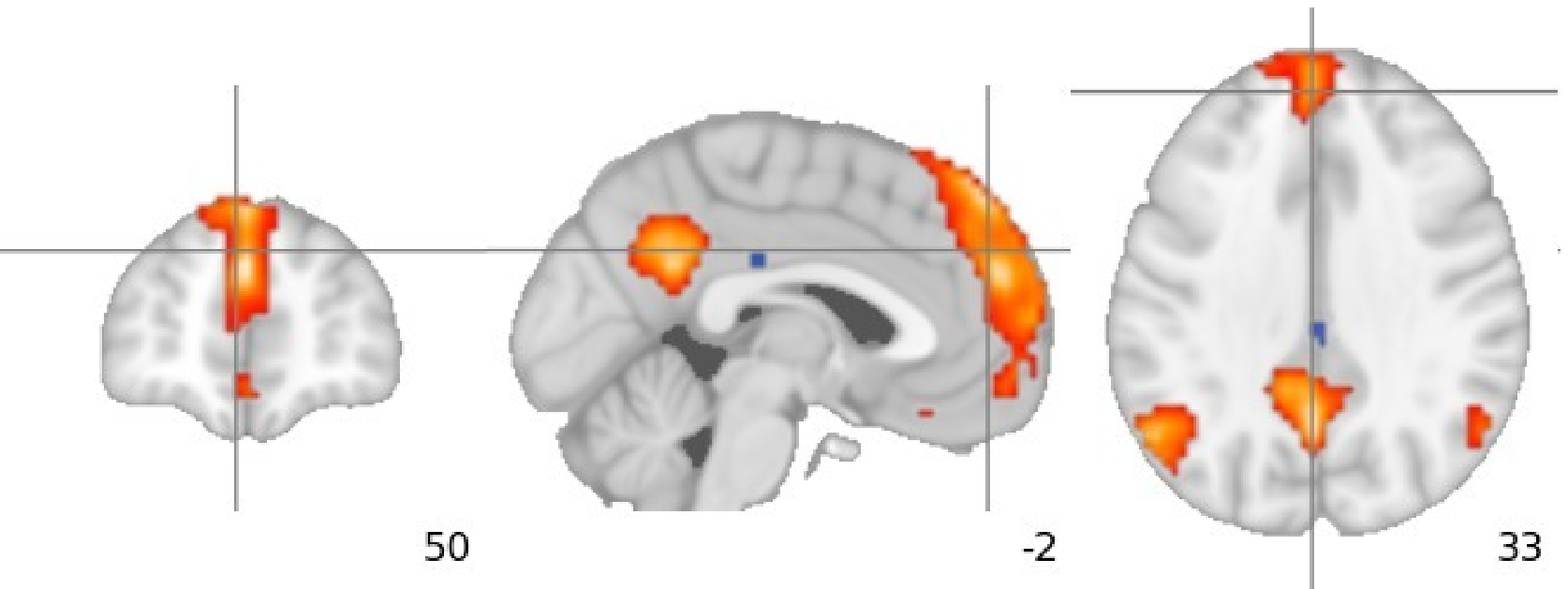}%
    \llap{\bfseries\sffamily\small (a) \hspace*{.62\linewidth}}%
}

\centerline{%
    \includegraphics[width=0.7\linewidth]{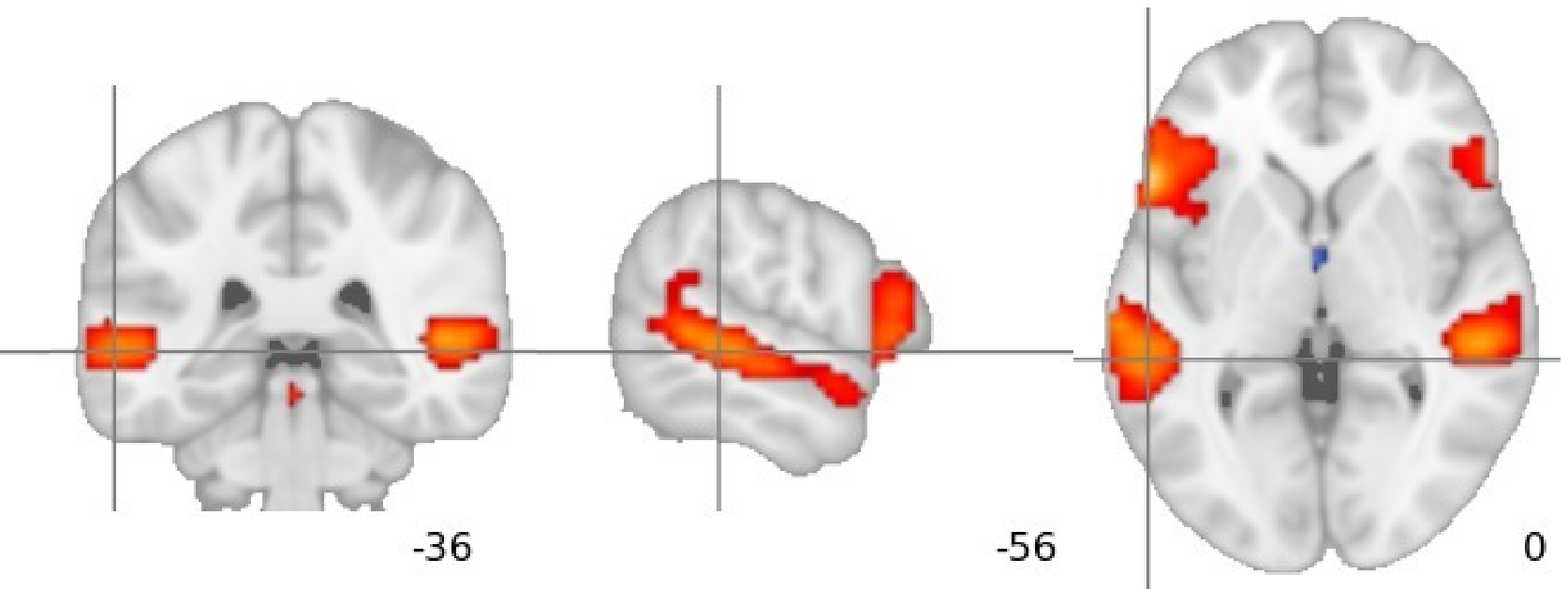}%
    \llap{\bfseries\sffamily\small (b) \hspace*{.62\linewidth}}%
}
\caption{
Intrinsic functional networks extracted from fMRI datasets of on-going activity,
adapted from \cite{varoquaux2010}. 
{\bfseries\sffamily (a)} Default mode network \cite{raichle2001}.
{\bfseries\sffamily (b)} Language network.
\label{fig:resting_state_networks}
}%
\end{figure}

\paragraph{Intrinsic functional networks} 
The seminal work of \citet{biswal1995} showed that correlations in
\emph{resting-state} brain activity, \emph{i.e.} in the absence of task,
can be used to reveal brain systems dedicated to a cognitive function,
often called \emph{intrinsic functional networks}. 
Indeed, in the presence or in the absence of explicit cognitive task, a
large fraction of brain activity is due to \emph{on-going} functional
processes.
After \citet{shulman1997} used task-independent spatial correlations maps
to uncover a new and important functional network \cite{raichle2001}, analysis of
fluctuations in on-going activity has been used to systematically map the
large-scale intrinsic functional architecture of the brain
\cite{fox2007,biswal2010}.
Data-driven extraction of the main functional networks from resting-state
data has been shown to be stable across subjects
\cite{damoiseaux2006,varoquaux2010}.
\noindent 
These networks form a relevant description of the brain: they are
found alike in on-going activity and in task-driven experiments, and
stand out as landmark structures in meta-analyses carried out on
current functional protocols \cite{smith2009}; they provide relevant
descriptors for large scale pathologies such as neurodegenerative diseases
\cite{seeley2009}; and they reflect the underlying anatomical
connectivity \cite{greicius2009,vandenheuvel2009}.
An important aspect of these networks is that many correspond
directly to set of regions that are specifically recruited for certain
cognitive processes, including higher-order functions, such as
language (Fig.~\ref{fig:resting_state_networks}.b) or executive
function \cite{seeley2007}. 
This decomposition of on-going activity
offers a insightful view on the functional architecture of the brain,
that is the interplay of a set of weakly-overlapping,
functionally-specialized, brain networks: \emph{``the human brain is
intrinsically organized into dynamic, anticorrelated functional
networks''} \cite{fox2005}.

\paragraph{Criticism of small-world analyses}
The analysis of brain architecture in terms of its graphical 
properties or its intrinsic functional networks gives differing points of
view: on the one hand the distance on the connectivity graph between
any two nodes of the small-world brain is small; on the other hand the brain
appears as decomposed in functionally-specialized and weakly-overlapping
networks of regions. While these two descriptions are not mutually
exclusive, small-world analyses have drawn some criticism.
\citet{ioannides2007} criticizes the random-graph description of brain
networks, as it is based on a small number of graph metrics that give a
fairly unspecific characterization of the brain. He argues for a
\emph{``hierarchy of networks rather than the single, binary networks
that are currently in vogue''}. More fundamentally, Ioannides raises the
issue that these properties can easily be confounded by observation
noise: \emph{``[in M/EEG] the activation of few (or many) uncorrelated
generators will [...] produce a small-world topology in a network
computed from the raw signal topography''} \cite{ioannides2007}. 
Indeed, \citet{deuker2009} find an inter-class correlation of less than
0.5 on the small-worldness index in a test-retest study using MEG for 2.5
minutes\footnote{2.5 minutes of acquisition time may seem short, yet for
MEG, it provides orders of magnitude more observations than a standard fMRI
experimental run} of resting-state acquisition. In addition,
simulation-based studies show that an observed small-world structure for
a graph empirically derived from time-series may solely be explained by
sampling effects \cite{bialonski2010,gerhard2011}.

It is thus legitimate to ask whether the small-world properties of
functional connectivity are indeed necessary to give a good description
brain activity, or whether a picture of brain function in terms of
separated cognitive networks is sufficient in view of the data.

\section{Statistical tools for connectivity analysis: Markov and Gaussian
graphical models}

\label{sec:statistical_tools}

\paragraph{Notations}
To clarify the mathematical presentation, we apply the following
conventions.
We write sets with capital letters, matrices
with  capital bold letters, $\B{A} \in \mathbb{R}^{n\times n}$, and we
denote $\|\B{A}\|$ the operator norm, $\|\B{A}\|^2 = \sum_{i, j = 1}^n
\B{A}_{ij}^2$, and $\|\B{A}\|_1$ the $\ell_1$ norm, $\|\B{A}\|_1 =
\sum_{i, j = 1}^n |\B{A}_{ij}|$. $\B{I}$ is the identity matrix.
Quantities estimated from the data at hand are written $\widehat{\B{A}}$.
$\B{A}^{-1}$ is the matrix inverse of $\B{A}$, $\B{A}^T$ is the
transposed matrix. Finally, we write $\B{A} \succ 0$, for a symmetric matrix
$\B{A}$ with strictly positive eigenvalues, \emph{i.e} a positive definite
matrix.

\subsection{Probabilistic modeling of brain network structure}

A limitation of the commonly used graph-theoretical descriptions of
functional connectivity is that they do not form a probabilistic model
of the functional signal. As such, they convey no natural notion of
goodness of fit and hypothesis testing of the graph properties in the
presence of noise is ill-defined.
This is why we resort to the analysis of graphical models specifying
a probability distribution for the signal. Namely we use Markov
models, that we apply to the functional-connectivity correlation
matrices in the framework of Gaussian graphical models.

\paragraph{Markov models: conditional independence graphs}
Measures of functional connectivity are based on correlations in the BOLD
signal. They cannot be easily interpreted in terms of transfer of
information or effective connectivity, that is \emph{the influence that a
neural system exerts over another} \cite{friston1994b}. 
Indeed, BOLD is a
very indirect measure of neural firing, the MRI signal observed contains
many non BOLD-related confounds, and correlation is a rough summary of
the dependence between two variables revealing, amongst other things,
indirect effects.

For these reasons, we focus our study on recovering the independence
structure of the observed functional signals. With the data at hand, fMRI
recordings with limited observations available, this is a challenging
statistical estimation problem. We use a class of probabilistic models
called \emph{Markov models} that specify the independence structure of
the variables that it describes. It can be represented as a graph, in which
case a node is statistically independent from nodes to which it is not
directly connected, conditionally on its neighbors.

In the context of brain functional connectivity, we are interested in
inferring the graphical structure from the correlations in the
observed activity. We frame this problem as estimating the Markov
structure\footnote{More specifically, we learn a undirected graph of
conditional independence. Such a model is often called a Markov Random
Field model.} of a Gaussian graphical model of brain activity.

\paragraph{Gaussian graphical models}
The study of functional connectivity focuses on second-order statistics of 
the signal, in other word Gaussian statistics. While the underlying
signals are most-likely not Gaussian, these measures have been shown
to capture well the independence structure underlying fMRI signals
\cite{smith2011}.
A multivariate Gaussian model with a specified Markov structure is called
a \emph{Gaussian graphical model} \cite{lauritzen1996}. We apply this 
class of models to the study of brain connectivity.

A centered Gaussian model is fully specified by the inverse of its
covariance matrix, known as its \emph{precision} matrix $\B{K}$, that
closely relates to conditional correlations between its variables. The
conditional independence between variables in the model are given by the
zeros in this matrix. Therefore the statistical estimation task that we
are interested in, is to find these zeros in the precision matrix.

We start from a brain activation dataset $\B{X} \in \mathbb{R}^{p \times
n}$ with $p$ variables, the activation of $p$ different brain regions, and
$n$ samples. We are interested in inferring a large-scale connectivity
graph, that is between many nodes, from a comparatively small amount of
observations. In these settings, estimating the covariance matrix, or the
precision matrix, from the data is an ill-posed problem (see for instance
\cite{ledoit2004,bickel2008}). First, if $n < \frac{1}{2}p(p+1)$, the number of
unknown parameters is greater than the number of samples. Second, the
various parameters to estimate are not mutually independent, as the
estimated covariance has the constraint of being positive definite.

Thus, the Markov structure of the data cannot be estimated simply by
thresholding the precision matrix. We resort to modern covariance
estimation tools, namely \emph{covariance selection}, as detailed below.
In this context, specifying a Markov structure acts as a regularization 
for the covariance matrix estimation: it injects a prior information that
a certain number of coefficients of the precision matrix are zero, and
thus decreases the number of coefficients to be estimated.

\citet{smith2011} have shown on realistic simulations of brain functional
connectivity that sparse $\ell_1$-penalized inverse covariance estimators
performed well to recover the graphical structure from noisy data.
\citet{huang2009} and \citet{varoquaux2010c} used such a sparse
covariance estimation procedure to infer the conditional independence
structure of a Gaussian graphical model on full brain data. 

\paragraph{Assessing model fit}
As a Gaussian graphical model defines a probability of observing data,
its log-likelihood can be used as a natural metric for assessing the 
goodness of fit of the model for some test data $\B{X}$: for a model
specified by its precision matrix $\B{K}$,
\begin{equation}
    \mathcal{L}(\B{X}|\B{K}) = \frac{1}{2} \left( \log \det \B{K} -\text{tr}\, 
    \B{K} \,\hat{\B{\Sigma}}_\text{emp}. \right) + \text{cst}
    \label{eq:model_likelihood}
\end{equation}
where $\hat{\B{\Sigma}}_\text{emp}$ is the empirical covariance of the
data: $\hat{\B{\Sigma}}_\text{emp} = \frac{1}{n} \B{X} \,\B{X}^T$.

For our application, an important point is to separate properties of
the functional connectivity networks extracted from the signal from
features learned on observation noise particular to the experimental
run. This problem is well known in statistics under the term of
\emph{overfit}. We use a standard technique to control overfit known
as \emph{cross-validation}: having learned a probabilistic model on a
given set of experiment run, we test its goodness of fit on unseen
data. Features learned by chance do not generalize to unseen data, as
it is independent from the training data.

\subsection{Covariance selection procedures}

As proposed by \citet{Dempster1972}, learning or setting conditional
independence between variables can be used to improve the conditioning of
an estimation problem, a technique referred to as \emph{covariance
selection}. Improved covariance estimation can therefore be achieved by 
estimating from the data a precision matrix with a sparse support, i.e., a 
small number of non-zero coefficients.

Selecting the non-zero coefficients to optimize the likelihood of the
model given the data (eq.~\ref{eq:model_likelihood}) is a difficult
combinatorial optimization problem. It is NP hard in the number of edges
and it therefore cannot be solved on moderately large graphs. To tackle
such problems efficiently, two strategies coexist: convex relaxation
approaches that lead to optimal solutions for a new problem that is a
convex approximation of the initial problem and greedy approaches that find
sub-optimal solutions.

\paragraph{Sparse penalized inverse covariance estimation}
A constraint based on the number of non-zero coefficients
leads to NP hard problems. It can be modified by replacing it by a
constraint on the $\ell_1$ norm of the solution. This leads to a
penalized estimator taking the form of the
following convex optimization problem \cite{Friedman2008}:
\begin{equation}
    \hat{\B{K}}_1 = \underset{{\B{K} \succ 0}}{\text{argmax}} \big(
	\log \det \B{K} -\text{tr}\,
	\B{K} \, \hat{\B{\Sigma}}_\text{emp}
    	- \lambda \|\B{K}\|_1 \big),
    \label{eq:penalized_estimation}
\end{equation}
where $\|\cdot\|_1$ is the element-wise
$\ell_1$ norm of the off-diagonal coefficients in the matrix. This
problem can be solved very efficiently in $\OO{p^3}$ time
\cite{banerjee2006, duchi2008}. 
This technique makes no assumption on the topology of the graph other than
its sparsity. Its major limitation is that the $\ell_1$ penalty may
bias the coefficients of the precision matrix.
It was employed successful to learn functional connectivity
graphs by \citet{huang2009} and \citet{varoquaux2010c} on full-brain fMRI
data, as well as \citet{smith2011} on realistic simulations of neural
interactions.

\paragraph{PC-DAG}
A greedy approach to learning the Markov independence structure of
observed data is given by the PC algorithm \cite{spirtes2001}. It
consists in pruning edges of a graph by successively testing for
conditional independence. Although it does not solve a global
optimization criteria, this algorithm is proved to be consistent
\cite{spirtes2001,robins2003}.
In the case of covariance selection, it can be used efficiently to
learn a sparse precision matrix, as in the \emph{PC-DAG} algorithm
\cite{rutimann2009}: the PC algorithm is applied to estimate a Markov
structure for the Gaussian graphical model, which is then used to
estimate the precision matrix. An important point is that the Markov
independence structure estimated is constructed as the \emph{moral 
graph}\footnote{A moral graph of a directed graph is the equivalent undirected
graph in terms of non-directed conditional independence relations
\cite{lauritzen1996}} of the (possibly) directed conditional independence
relation extracted by the PC algorithm. This graph is limited by the PC
algorithm to a small node degree. This implies that the graphs extracted
by the PC-DAG are unlikely to contain large cycles.
This procedure is interesting because its computational cost scales
exponentially with the maximum number of neighbors on the underlying
graph, rather than the number of possible edges.


\paragraph{Shrinkage estimates}
Finally, a more naive but quite efficient approach in practice consists in 
regularizing the estimate of the precision matrix by adding a diagonal matrix to the empirical covariance before computing its inverse. It amounts to an $\ell_2$ shrinkage by penalizing uniformly off-diagonal terms:
\begin{equation}
    \hat{\B{K}}_2 = ( \hat{\B{\Sigma}}_\text{emp} + \lambda \,\B{I})^{-1}
\label{eqn:k_diagonal}
\end{equation}
\citet{ledoit2004} have introduced a closed formula 
for a choice of $\lambda$ leading to a good bias-variance
compromise. Empirical results show that this method can achieve 
good generalization scores even with small sample sizes. The intuition
behind this good performance is that a sample covariance matrix
tends to over-estimate pair-wise correlations in the small-sample 
limit\footnote{Indeed the plug-in estimator for pair-wise correlations,
\mbox{$\hat{r} = \langle\B{x}, \B{y}\rangle / (\|\B{x}\| \, \|\B{y}\| )$},
which also corresponds to the scaling of off-diagonal coefficients in the 
empirical covariance, is known to be biased at small samples. In
particular, this bias implies that the most probably observed correlation
is smaller than the population value, as discussed in 
\cite{fisher1915} page 520.}.
Shrinking them to zero thus improves the estimate.
This procedure gives a good baseline to compare goodness of fit of
models, 
however, it does not extract a graph structure from the observed
time series, as it does not set zeros in the
precision matrix. 

\subsection{Decomposable graphical models}

Here, we introduce the notion of \emph{decomposable graphs},
and present some related properties interesting to describe brain
functional networks.

\begin{figure}
\hfill%
\centerline{%
    \includegraphics[width=.8\linewidth]{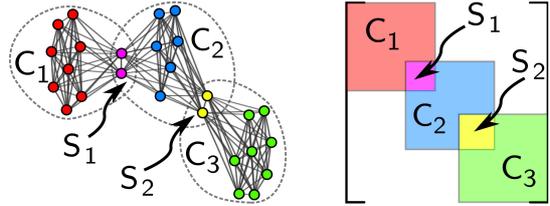}%
}

\caption{Structure of a decomposable graphical
model. 
{\bf Left:} graph representation. {\bf Right:} corresponding 
precision matrix.
Nodes can be ordered in such a way that they form a succession of
cliques $C_i$, that are interconnected solely by separating sets $S_i$.
\label{fig:decomposable_graph}
}%
\end{figure}

A graph is said to be {\sl chordal} if each of its cycles of four or
more nodes has a chord, i.e., an edge joining two nodes that are not
adjacent in the cycle. A chordal graph can be divided in an ordered
sequence of cliques\footnote{%
A clique is a fully-connected sub-graph: each nodes is connected to every
other node.
}
${C_i}$ \cite{lauritzen1996}, as illustrated in
Fig.\,\ref{fig:decomposable_graph}. The intersection
between two successive cliques $C_i$ and $C_{i+1}$ forms a complete
subgraph\footnote{A complete graph is a graph in which all pairs
  of nodes are connected.} called a separating set $S_i$.  When the
structure of a graphical probabilistic model is chordal, the model is said 
to be {\sl decomposable} \cite{dawid1993}. 
The different cliques $C_i$ are then mutually
independent conditionally on the separating set $S_i$, and
the likelihood factors in terms corresponding to the different cliques and 
separating sets:
\begin{equation}
    \mathcal{L}(\B{X}|\B{K}) = \sum_{C_i} \mathcal{L}(\B{X}_{C_i}|\B{K}_{C_i}
)
    - \sum_{S_i} \mathcal{L}(\B{X}_{S_i}|\B{K}_{S_i}),
    \label{eq:likelihood_decomposable}
\end{equation}
where $\B{X}_{C_i}$ denotes the dataset reduced to the variables of clique
$C_i$ and $\B{K}_{C_i}$ denotes the corresponding reduced precision
matrix \cite{lauritzen1996}. While there exists no closed formula for the
Maximum Likelihood Estimate (MLE) of the precision matrix for an arbitrary 
sparse graphical Gaussian model, the MLE for a decomposable model can be
written as:
\begin{equation}
    \B{K}_{MLE} = \sum_{C_i} [\hat{\B{K}}_{C_i}]^0
    - \sum_{S_i} [\hat{\B{K}}_{S_i}]^0,
    \label{eqn:k_decomposable}
\end{equation}
where $[\hat{\B{K}}_{C_i}]^0$ denotes the MLE for the precision matrix of
the fully-connected subset of $\B{X}$ reduced to the clique $C_i$, that
is the inverse empirical covariance matrix, padded with zeros to the size
of the full model \cite{lauritzen1996}.
Identifying the cliques and separating sets in a graphical model opens the
door to an efficient estimation of the maximum-likelihood precision
matrix. Note that a non-decomposable model can be turned in a
decomposable model: cordless cliques can be completed to turn the graph
in a chordal graph.

The motivation for using decomposable models to learn brain
connectivity is twofold. First, the covariance matrices are estimated
at the clique level, which are smaller problems. This strongly limits
the estimation error. More importantly, the description of the brain
that decomposable models give, cliques of regions that are mutually
independent conditionally to set of hubs nodes, seems to match well
the intuition of the functional networks composing brain architecture.
Models sharing a similar intuition but not relying on a graphical
description have been popularized in neuroimaging through the concept of
\textit{integration} of brain regions within a network \cite{marrelec2008}.

\section{Efficient learning of decomposable structures}

\label{sec:algorithm}

In this section, we introduce a new algorithm to learn decomposable
models on large covariance matrices without a priori knowledge of the
relevant cliques. Indeed, to date, none of the standard covariance
selection procedures for large-scale graphs seeks a decomposable model. 
While Bayesian approaches
to learning Gaussian graphical models often rely on decomposable models,
they require sampling of the space of models, and therefore are
intractable on problems with more than a few dozen of nodes
\cite{giudici1999,donnet2010}. \citet{marlin2009} recently proposed an
algorithm to impose a clique structure on Gaussian graphical models.
However, their approach, with an $\OO{p^4}$ complexity, requires a known
ordering of the nodes.

We propose a fast algorithm, that we call {\sl FastDecomp}, to learn a
decomposable graph with a large-scale structure common to several
datasets. Our goal is to find a structure of conditional independence
between successive blocks of variables. Our algorithm is based on a
greedy approach to select non-independent variables.

\paragraph{The FastDecomp algorithm}
Similarly to the PC-Algorithm, we start from a complete graph
$\mathcal{G}$, compute pair-wise partial correlations conditioning on all
other variables, and use them to test conditional independence between
variable pairs to prune the graph. However, to keep bounds on the
computational cost, we do not pursue to test conditional independence
exhaustively. To test for independence between
nodes $\B{X}_i$ and $\B{X}_j$, conditioning on 
$\B{X}_K, K = \{1 \dots p\} \backslash \{i, j\}$, 
we first test conditional independence using Fisher's z-transform of the
estimated partial correlation\footnote{We use the
Ledoit-Wolf estimator to estimate the partial correlation, as it gives a
good bias-variance compromise.}:
\begin{equation}
    z (i, j | K) = \frac{1}{2} \log \biggl( 
    \frac{1 + \hat{\rho}_{i, j | K}}{1
    - \hat{\rho}_{i, j | K}} \biggr).
\end{equation}
$z (i, j | K)$ has approximately a $\mathcal{N}(0, (n - p - 1)^{-1})$
distribution if $\rho_{i, j | K} = 0$ (see \citet{rutimann2009} for
similar considerations). 

\begin{algorithm}[tb]
   \caption{FastDecomp.
RCM stands for Reverse Cuthill-McKee algorithm \cite{cuthill1969}.
LedoitWolf stands for covariance estimation as in Eq.~\ref{eqn:k_diagonal}, 
\cite{ledoit2004}.}
   \label{alg:fastdecomp}
\begin{algorithmic}
   \STATE {\bfseries Input:} $\B{X}$, threshold $\beta$
   \STATE {\bfseries Output:} estimated precision matrix $\hat{\B{K}}$,
perfect elimination $order$
   \STATE Compute $\hat{\B{K}}_\text{LW} \leftarrow \bigl(\text{LedoitWolf}(\B{X})
		\bigr)^{-1}$
   \STATE Initialize $\mathcal{G}$ to a complete graph.
   \FOR{$i=1$ {\bfseries to} $p$}
     \FOR{$j=1$ {\bfseries to} $p$}
	\STATE Here we use  
	\vspace*{-1em}
	\[\hat{\rho}(i, j |
\{1 .. p\} / \{i, j\}) = 
\frac{(\hat{\B{K}}_\text{LW})_{i, j}}{
\sqrt{(\hat{\B{K}}_\text{LW})_{i, i}\,
(\hat{\B{K}}_\text{LW})_{j, j}}
},
\]
       $z (i, j | K) \leftarrow \hat{\rho}_{i, j | K}$
	\IF{$(z (i, j | \{1 .. p\}
	\backslash \{i, j\}) < \beta$} 
	    \STATE Remove edge $i, j$ from $\mathcal{G}$.
	\ENDIF
     \ENDFOR
   \ENDFOR
   \STATE $i_0, order = \text{RCM}(\mathcal{G})$
   \STATE Color the vertices of $\mathcal{G}$ with $order$
   \STATE Initialize $\hat{\B{K}}$ to a $p \times p$ matrix of zeros.
   \STATE Set $\B{X} \leftarrow [\B{X}^1, \dots, \B{X}]$.
   \STATE Sort $\B{X}$ according to $order$.
   \REPEAT
	\STATE $i_1 \leftarrow$ largest node in $adj(i_0)$ given $order$.
	\STATE $\hat{\B{K}} \leftarrow \hat{\B{K}} + \bigl[\bigl(\text{LedoitWolf}(\B{X}_{i_0\dots i_1})
		\bigr)^{-1}\bigr]^0$
	\IF{$i_1 \neq p$}
	   \STATE $i_0 \leftarrow$ smallest node in $adj(i_1 + 1)$ given $order$.
	   \STATE $\hat{\B{K}} \leftarrow \hat{\B{K}} - \bigl[\bigl(\text{LedoitWolf}(\B{X}_{i_0\dots i_1})
		\bigr)^{-1}\bigr]^0$
	\ENDIF
   \UNTIL $i_1 = p$
\end{algorithmic}
\end{algorithm}

Once all the pair-wise partial correlations are tested, we find a good 
completion of the resulting graph to a chordal graph\footnote{Finding the
best possible chordal completion is know to be NP-hard
\cite{papadimitriou1976}.}. For this we apply
the symmetric Reverse Cuthill-McKee (RCM) algorithm \cite{cuthill1969} 
on the corresponding
adjacency matrix to obtain a
peripheral node\footnote{A peripheral node of a graph is a node for
which the shortest path to another node is the diameter of the graph, that is
the maximum shortest path between all nodes. The
reverse Cuthill-McKee algorithm does not directly output a peripheral
node, but we use its graph traversal and dynamical programming to
find a pseudo peripheral node for little added cost.}
and an ordering minimizing the envelop of the graph. 
We start from the
peripheral node and jump successively to the adjacent node furthest
according to the RCM ordering, which gives us an enumeration of the
cliques $C_i$ of the chordal graph, and the separating sets $S_i$.

Finally, on each clique $C_i$ (resp. separating set $S_i$) we estimate
the precision matrix by applying the Ledoit-Wolf estimator to the
observed data, restricted to the clique (resp. separating set). We obtain
the final precision matrix using formula \ref{eqn:k_decomposable}. See
listing \ref{alg:fastdecomp} for a precise description of the algorithm.

\paragraph{Computational cost}
The Reverse Cuthill-McKee can be implemented in $\OO{p^2}$
\cite{chan1980}. The bottleneck of our algorithm is thus the last loop:
calculating the precision on each clique and each separating set. The
cost of this loop as bounded by $p$ times a matrix inversion. 
The {\sl FastDecomp} algorithm is thus bounded by
$\OO{p \times p^3}$.

\section{Experimental results: graphical models of on-going brain
activity}

\label{sec:experiments}

\subsection{A resting-state fMRI dataset}

We apply the various covariance-selection methods to estimate the Markov
independence structure from a resting-state fMRI dataset acquired in a
previous experiment \cite{sadaghiani2009}. 12 subjects were scanned in a
resting task, resulting in a set of 810 brain volumes per subject
acquired with a repetition of 1.5 second. The data were preprocessed
using the SPM5 software\footnote{Wellcome Department of Cognitive
Neurology; \url{www.fil.ion.ucl.ac.uk/spm}}, including in particular the
spatial normalization to standard template space. As in
\citet{achard2006}, or \citet{huang2009}, we extract brain-activation
time series on a parcellation of the gray matter in 105 non-overlapping
regions\footnote{The complete atlas comprises 110 regions, but as the
field of view of our fMRI images is reduced, it does not cover the
cerebellum.}. As we are interested in modeling only gray-matter correlations,
we regress out confound effects obtained by extracting signals in the
white matter, the cortico-spinal fluid regions, as well as the rigid-body
motion time courses estimated during data pre-processing.

\begin{figure}
\includegraphics[width=\linewidth]{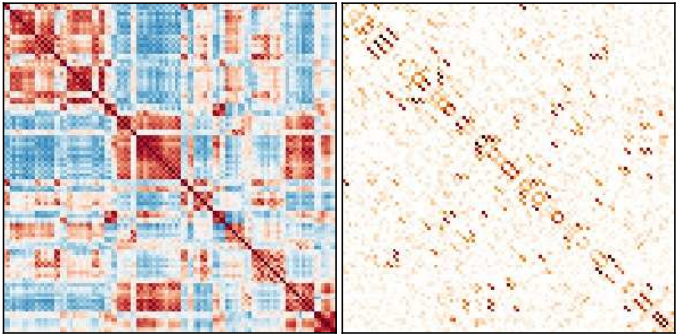}
\caption{Empirical covariance and associated empirical
precision for one subject. On the precision
matrix, the values on the diagonal are not represented. 
}
\label{fig:subject_emp_covs}
\end{figure}

The correlation matrix of the resulting signals can be seen in Figure
\ref{fig:subject_emp_covs} (left). On this figure, regions are ordered
according to the labels of the atlas that was used to define them
\cite{Tzourio-Mazoyer2002a}. This ordering groups together regions that
belong to the same general bilateral anatomical structures. A first
visual analysis reveals large blocks of correlated regions. This can be
interpreted as the signature of the so-called cognitive networks. 

The corresponding empirical precision matrix (Figure
\ref{fig:subject_emp_covs}, right) reveals the partial correlations
between regions. It appears more sparse; in particular the
block-structure is not visible.

\subsection{Generalization performance of the various estimators}

We learn from this dataset different Gaussian graphical models of brain
activity using the different covariance selection procedures: sparse
penalized inverse covariance estimation\footnote{We use an
implementation of the $\ell_1$-penalized inverse covariance estimator
following \citet{Rothman2008}}, PC-DAG, and FastDecomp. We evaluate the
goodness of fit of the different models using a leave-one-out procedure:
we learn a sparse precision on 11 subjects\footnote{We apply the
estimators on the concatenated individual data: $\B{X} = [\B{X}^1, \dots
, \B{X}^S]$, after detrending, band-pass filtering, and
variance-normalizing the individual time series. The generative model underlying
this concatenation corresponds to the hypothesis that all the different
observations are drawn from the same distribution. It is a common
assumption in unsupervised analysis of resting-state dataset, popularized
with ICA \cite{calhoun2001a}.} and test the likelihood of the data
corresponding to the $12^{th}$ subject in the model described by this
precision matrix. In the likelihood term (eq. \ref{eq:model_likelihood}),
we use correlation matrices, rather than covariances, as we are not
interested in the variance terms, but only in the correlation structure. 

As a baseline for the generalization likelihood, we use the Ledoit-Wolf
diagonal shrinkage estimator\footnote{We also tried setting the amount of
shrinkage by nested cross-validation, rather than using the Ledoit-Wolf
oracle. We found that it gave the same generalization score.}, that does
not impose any structure or sparsity on the model. Each algorithm has a
parameter that controls the degree of sparsity of the estimated graph:
for the sparse penalized estimator it is the amount of $\ell_1$
penalization --$\lambda$ in equation \ref{eq:penalized_estimation}-- and
for the greedy methods, PC-DAG and FastDecomp, it is the threshold on the
conditional independence test used to build the Markov independence graph
-- $\beta$ in listing \ref{alg:fastdecomp}. If we set these parameters to
maximize the likelihood of left-out data, we find that the best
performing solutions are achieved with very little sparsity or
penalization.

\begin{figure}
\centerline{%
    \includegraphics[width=.8\linewidth]{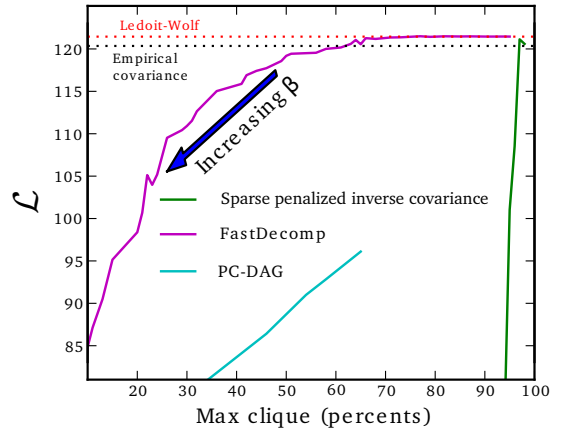}%
}
\caption{
Cross-validation scores of the sparse precision estimators
compared as a function of the width of maximum clique that they yield. 
For the FastDecomp algorithm, this width is varied by setting the $\beta$
parameter of the algorithm.
\label{fig:cv_scores}
}
\end{figure} 

The PC-DAG approach is fast when underlying graphs are very sparse.
However, to achieve good cross-validation scores on our data, the p-value
on the conditional independence test of the PC algorithm should be very
small. As a result, the algorithm explores denser graphs. As the
computational cost is in $\OO{e^k}$, where $k$ is the maximum node degree
of the graph, we could not reach an optimal point in reasonable time. We
interpret this result by the fact that the brain-connectivity network is
poorly represented by a network with small maximum node degree. 

\subsection{Trading off goodness of fit for decomposition into networks}

We are interested in interpreting the graphs in terms of functional networks.
As the networks found can be overlapping, the total number of cliques in
the decomposable graph is not a good indication of the amount of
``decomposition''. For this purpose, we study the maximum clique size,
which correspond to the maximum number of nodes in a functional network  (Figure
\ref{fig:cv_scores}). With a small maximum clique size, the likelihood of
the model (Eq. \ref{eq:likelihood_decomposable}) factors in terms
comprising a small amount of brain regions, and the decomposition into
conditionally-independent networks in more easily interpretable. Note
that, in this case, the description of networks is underpinned by a
probabilistic model in which they appear in independent terms, unlike
when using ICA
\cite{calhoun2001a,kiviniemi2003,beckmann2005,varoquaux2010}, or
seed-based correlation analysis \cite{biswal1995,cordes2000,fox2007}.

\begin{figure}
\centerline{%
    \includegraphics[width=.8\linewidth]{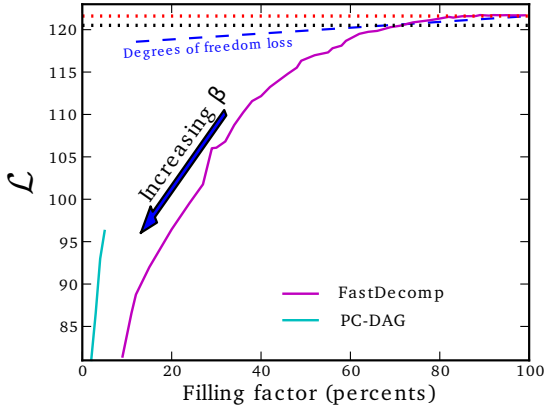}%
}
\caption{
Cross-validation scores as a function of the filling factor of
the graph: the ratio of edges to the total possible number of edges, $p^2$.
The dashed blue
line gives the loss in likelihood explained by the loss in degrees of
freedom of the model compared to the Ledoit-Wolf model  as the
filling factor of the precision matrix decreases.
\label{fig:cv_scores_filling}
}
\end{figure} 

\paragraph{Likelihood as a function of clique size} 
Decomposable models estimated using FastDecomp maintain a generalization
score (likelihood on left out data) as good as the baseline for maximum
clique size as large as 70\% of the number of nodes but for smaller
cliques their score decreases, and starts dropping quickly for cliques
smaller than 30\% (Figure \ref{fig:cv_scores}).
While the models estimated with the sparse penalized inverse covariance
estimator or the PC-DAG algorithm are not decomposable, the graph can be
completed into a chordal graph\footnote{This procedure is often called
triangularization of the graph.} using the RCM algorithm. Increasing the
penalization to create sparse graphs with the penalized estimator leads
to a important loss in generalization scores without a sizable clique
width reduction (Figure \ref{fig:cv_scores}). Indeed, completing the
graphs to be chordal creates large cliques unless they are very sparse.
This high sparsity is enforced by using a large penalization, which
introduces bias in the estimated coefficient and is detrimental to the
generalization likelihood. The PC-DAG does not suffer from this
limitation. However, we could not use it to explore graphs with a degree
higher than 14 due to available computational power as the complexity of
this algorithm is exponential in the maximum degree. Although very sparse, these
graphs already have large cliques.

\paragraph{Selecting a decomposed model}
As the maximum clique size decreases, so does the filling factor of the
precision matrix. As a result, we are comparing models with
different degrees of freedom and we should account for the difference in
our model comparison, based on a likelihood ratio test. The expected
difference in log-likelihood, under the null hypothesis that two models
fit as well the data, is given by half the difference in degrees of
freedom. When accounting for this loss (Figure \ref{fig:cv_scores_filling}),
we find that for filling factors above 70\%, which
corresponds to a maximum clique width of 60\%, the model learned by
FastDecomp performs as well as the baseline non-sparse model.

\subsection{Structure of the models}

\begin{figure}
\begin{minipage}{.5\linewidth}
    \hspace*{-1ex}%
    \includegraphics[width=\linewidth]{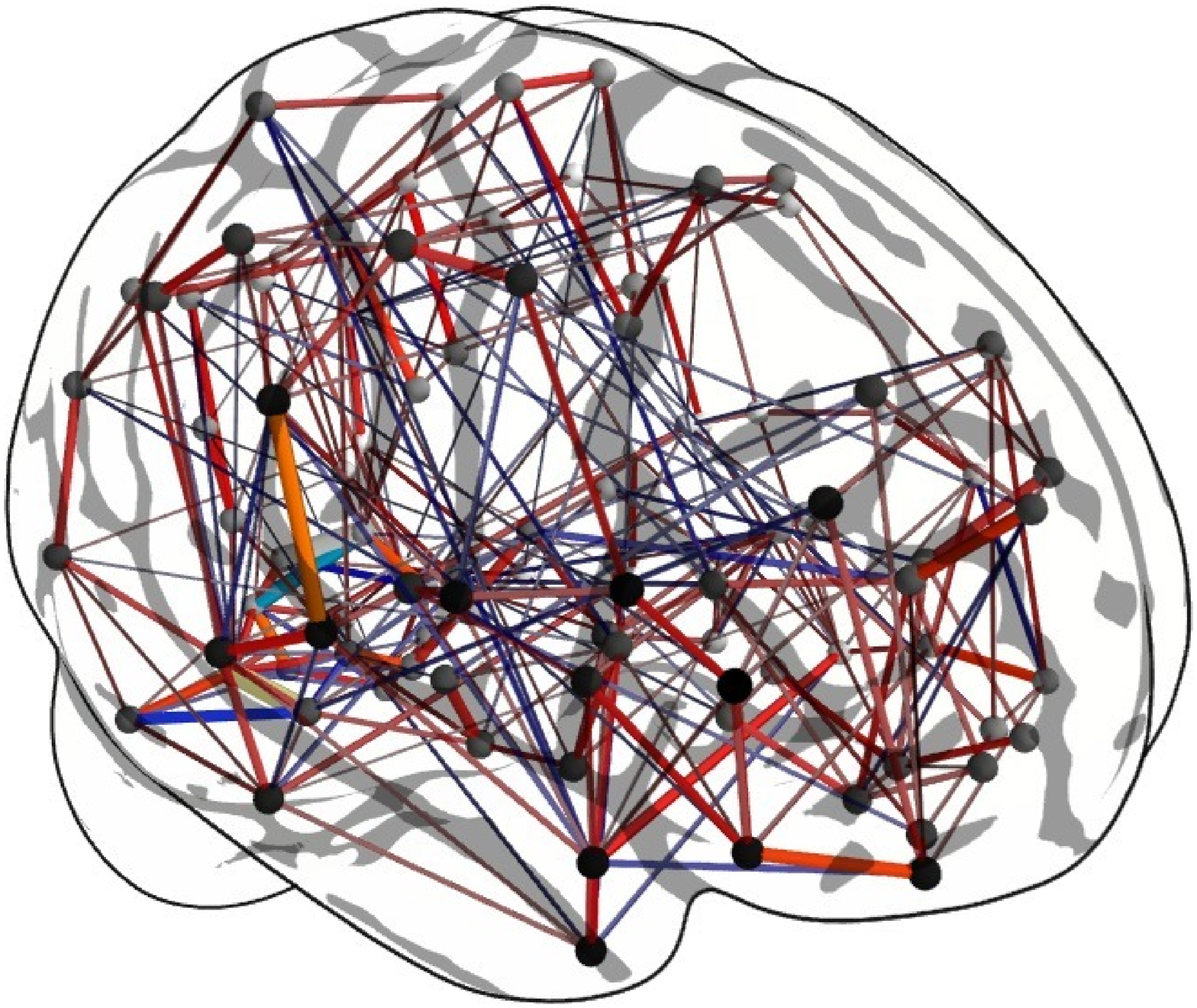}%
\end{minipage}%
\begin{minipage}{.5\linewidth}
    \hspace*{1ex}%
    \includegraphics[width=\linewidth]{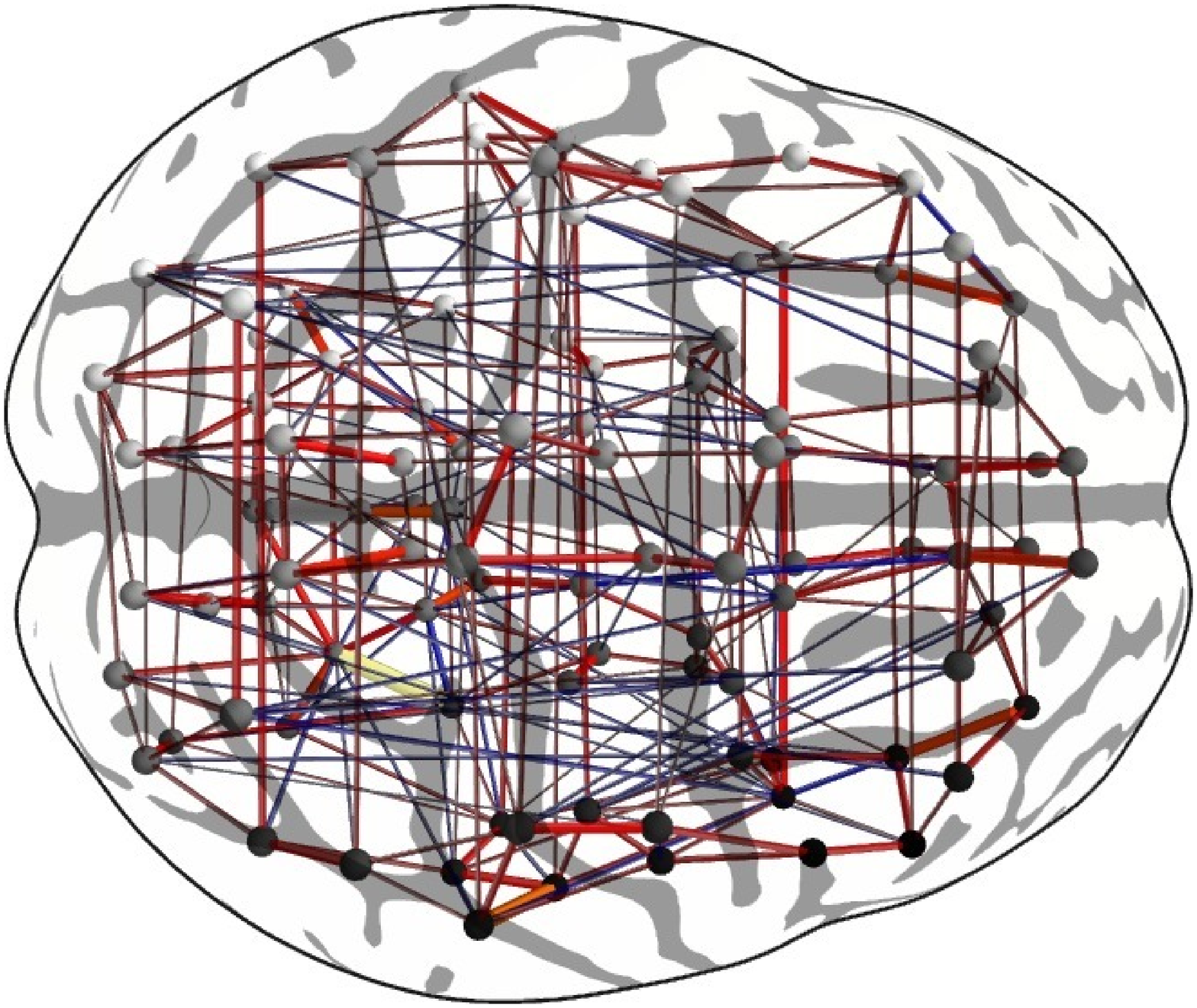}%
    \hspace*{-1ex}%
\end{minipage}
\caption{Graphical model of brain connectivity estimated by the PC-DAG
algorithm. The graph estimated is very sparse, due to cost of the
algorithm, exponential in the maximum node degree. However, it displays
some clear structure that recalls the structure of small-world graphs 
described by
\citep{watts1998}. First, neighboring nodes on the cortical surface are
connected, in a 2D lattice-like structure. Second, connections outside
the lattice structure create \emph{shortcuts} in the graph:
opposite homologous regions, as well as a few inter-hemispheric
connections.}
\label{fig:pc_dag_brain}
\end{figure}

Here we discuss the structure of brain connectivity that appears in the
graph learned.

\paragraph{Very sparse models}
Only very sparse graph structures are directly interpretable. As can be
seen on figure \ref{fig:cv_scores_filling}, PC-DAG is the algorithm best
suited for this purpose as, unlike penalized methods, it does not bias the
estimation of the precision matrix coefficient, and, unlike the FastDecomp
algorithm, it does not force small cliques. It is interesting to note that
these graphs have large cliques (Fig \ref{fig:cv_scores}), even though we
were limited to exploring small maximal node degrees\footnote{Strictly
speaking, this node degree corresponds to the underlying DAG computed by
the PC-DAG algorithm, and not the final undirected graph.} due to
available computational power. Such graphs (Figure
\ref{fig:pc_dag_brain}) have a lattice-like structure --nearest-neighbor
connections on the cortex-- with the addition of inter-hemispheric
homologous connections and a few long-distance connections.

\begin{figure}[p]
\includegraphics[width=\columnwidth]{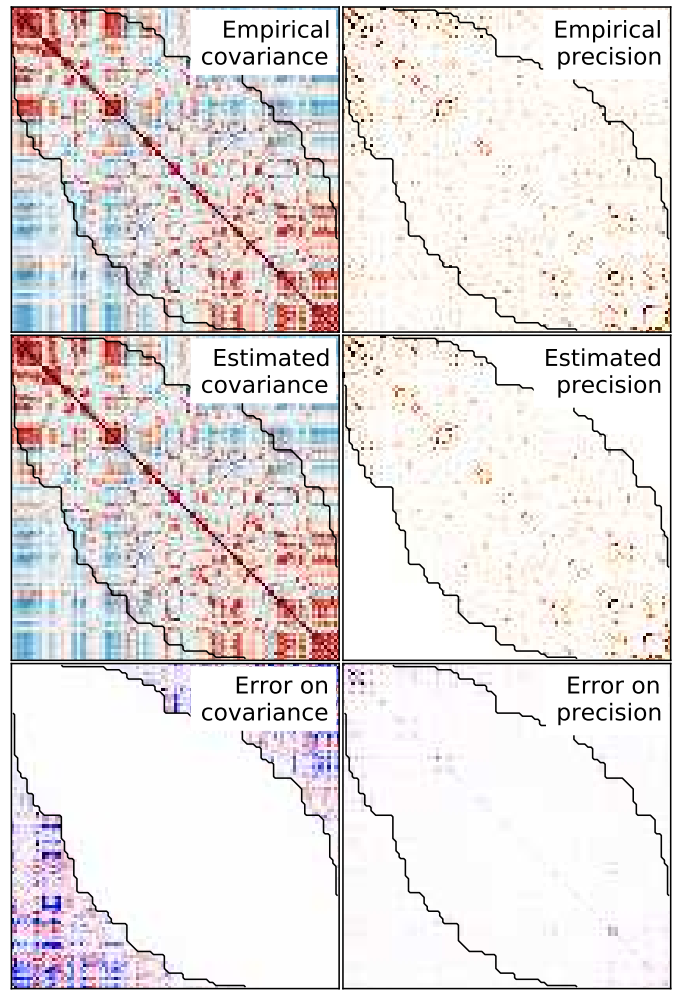}
\vskip -0.1in
\caption{Reordered empirical covariance and precision matrix,
as well as estimates by FastDecomp, and the difference between empirical
and estimated values. The parameter $\beta$  was set to limit
the size of the largest clique to 50\% of the nodes.
Compared to figure \ref{fig:subject_emp_covs}, the nodes have been reordered in
a complete elimination ordering of the decomposable model.
The outline of the cliques is drawn using a black line. Note that, while
the empirical precision and the estimated precision seem to differ
significantly due to the values set to zeros outside the cliques
selected, the empirical and estimated covariance look very similar.
The errors on the covariance matrix inside the cliques selected are not
visible, as they are 1/100 times smaller than outside the cliques.
}
\label{fig:decomposed_model}
\vskip -0.2in
\end{figure}

\begin{figure}[p]
\vskip 0.2in
\hspace*{-.05\columnwidth}%
\includegraphics[width=1.1\columnwidth]{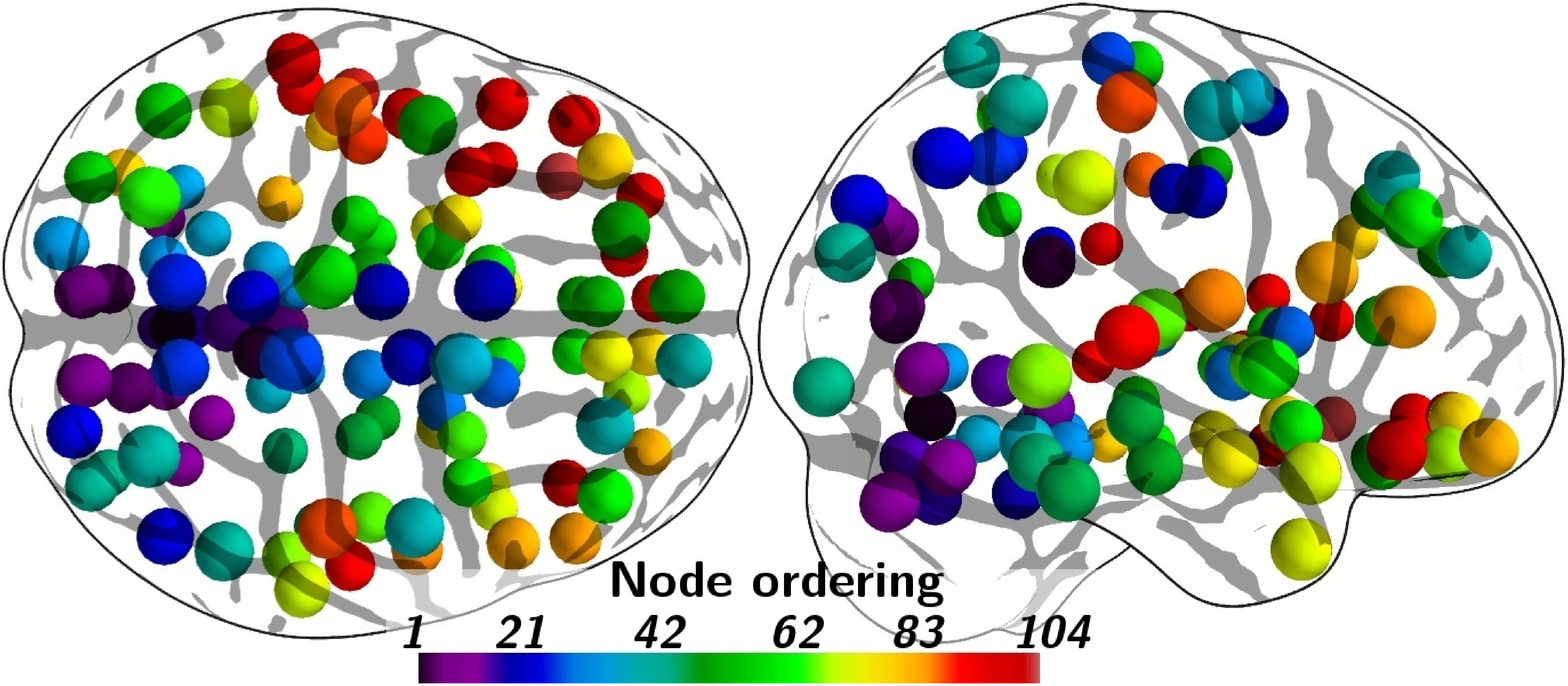}%
\vskip -0.1in
\caption{Node ordering in the decomposable model, represented on the region 
centers, in a view of the brain. The color of the nodes indicates the
order in which they appear, in the decomposable model estimated in Fig.}
\ref{fig:decomposed_model}
\label{fig:3D_brain}
\vskip -0.2in
\end{figure}

\paragraph{Decomposition in networks}
To assert the meaning of the decomposable models in neuroscientific
terms, we apply FastDecomp with $\beta$ chosen such that the maximum
clique width is 50\% of the full graph. As highlighted by the above
cross-validation study, this results in a small loss in generalization
scores. However, the resulting cliques and separating
sets form small groups of regions and are therefore more interpretable. The
estimated banded precision matrix is given in
Figure~\ref{fig:decomposed_model}. The differences between empirical and
estimated covariance accumulate mainly outside the cliques that were
selected by FastDecomp, corresponding to the long-range interactions between 
networks far apart according to the ordering learned.
In Figure~\ref{fig:3D_brain}, we
display the regions forming the nodes of the graph, colored by their 
node ordering on a standard brain model. 

It is clear that the ordering is not random: regions of similar function
are grouped together. For instance, the purple nodes form the primary
visual cortex and are connected to blue regions along the well-known
dorsal visual pathway, going up to the superior parietal lobule
\cite{goodale1996}.
The separation of brain structure from the outer surface versus cingular
regions, located in between the hemispheres, is also striking, as well as
the symmetry of the model.
However, more interesting is that the red regions, representing high node
order, form a left-lateral network in the temporal lobe. It resembles the
language network, the lateralized network most often recruited that is
displayed on figure \ref{fig:resting_state_networks}b, as estimated from
the same resting-state dataset using an ICA-based method
\cite{varoquaux2010}.
We note that the ordering of regions located in the central cliques is
more difficult to interpret as it appears to be representing several 
cognitive networks. This is consistent with the fact that these regions are part
of larger cliques, that do not separate small structures.
We defer a more thorough neuroscientific interpretation of the present
results, and in particular a link with the knowledge available from
anatomical connectivity data, to future work. 


\paragraph{A high bandwidth structure}
While the graphical structures that we extract are sparse, we can
see that they must have a small average path length to fit the data.
Indeed, we find that the adjacency matrix of the graph must have a
high bandwidth, as the likelihood drops when forcing a small
bandwidth\footnote{The RCM is an algorithm commonly used for ordering
  a graph to evaluate its bandwidth.}  as for small clique sizes on
figure \ref{fig:cv_scores}. The graph is therefore a high-bandwidth
graph \cite{gross2004}, which implies it has a small
diameter\footnote{The diameter of a graph is the maximal value of the
  shortest path lengths for all pairs of nodes in the graph}
\cite{chinn1982}. This in turns implies that average path length on
the graph is small. In layman's term, any two nodes are always
connected by a small distance on the graph, as they can not be
separated by more than a few large networks.

\paragraph{Small word properties}
The high bandwidth that we observed is characteristic of small-world
graphs. Our analysis shows that this property is required for a model to
fit the data well. The other aspect of a small-world graph, local
properties of the graph such as clustering coefficient, is sensitive to
sampling noise or the choice of thresholding or any other
sparsity-enforcing parameter.	It appears that, with our dataset, little
sparsity is best to maximize goodness of fit\footnote{The small amount of
sparsity may be due to spurious effects, such as inter-subject
variability, or the definition of the regions used to extract the signal.
We used a standard parcellation of the brain based on the anatomy of a
single subject that is used routinely in small world analysis of fMRI --
to list only a few \cite{salvador2005,achard2006,huang2009}-- although it
has been shown that small-worldness, measured by the ratio of clustering
coefficient and average path length, varied as a function of the brain
parcellation used \cite{wang2009}.}. To study the local properties of the
graph, we can use the results of the PC-DAG algorithm. Indeed, the
chordal completion of the graph used in FastDecomp increases artificially
the clustering coefficient, and the penalized estimators give precision
matrices that are not sparse enough to be considered as small
world\footnote{These precision matrices have many small coefficients,
such that thresholding them would put give sparsity level adapted to
studying small world properties, however there is no
statistically-controlled way of thresholding, as it would also create non
positive definite matrices, for which the likelihood of the model is not
defined.}. On the graph extracted by the PC-DAG with the best goodness of
fit, we find an average shortest path length of 3.0 and an average node
clustering coefficient of 0.19. These numbers are characteristic of small
word graphs \cite{watts1998} and are consistent with the graph properties
reported with previous functional-connectivity analysis of fMRI that did
not rely on Markov models \cite{bassett2006}.

We note that the PC-DAG algorithm is ill suited to recover small-world
graphs, where the presence of hub nodes leads to high computational
costs. On the opposite, FastDecomp is well suited for extracting large
interpretable cliques on such data, but not the local properties.

\section{Conclusion}

We have applied different covariance selection procedures to learn
probabilistic graphical models --Markov models-- of brain
functional connectivity from resting-state fMRI data.
We introduce a definition of the large-scale functional networks as the
cliques of a decomposable graphical model.
To learn efficiently decomposable models on graphs presenting hub nodes,
such as small-world graphs, we have introduced a new algorithm.
By setting the width of the
largest clique of the graph, we have investigated the compromise between
models interpretable in terms of independent functional networks, and models 
that generalize well. 

We find that the brain is best represented by a high-bandwidth graph
that cannot be decomposed into small cliques: forcing a strongly
decomposable representation breaks its small-world
properties. However, we can learn a simplified model of the data in
which the variables are decomposed into smaller
conditionally-independent units that can be interpreted as functional
networks, matching the current knowledge in cognitive neuroscience. To
summarize adequately brain functional connectivity with segregated
networks, these must encompass many brain regions and be strongly
overlapping. These preliminary results provide further evidence that
the functional connectivity signals observed in fMRI reflect the
small-world properties of brain connectivity.  Markov models,
describing the conditional independence relations of the data, are a
promising tool to investigate the large scale architecture of the
brain via its functional connectivity.

\section*{Acknowledgments}

We wish to thank Andreas Kleinschmidt for many enlightening discussions
on the cognitive meaning of on-going activity as well as Sepideh
Sadiaghiani for acquiring the data used in the paper. We are also
grateful to the anonymous reviewers for helpful remarks. This project was
partly funded from an INRIA-INSERM collaboration. The fMRI set was
acquired in the context of the SPONTACT ANR project (ANR-07-NEURO-042).
Estimation algorithms were implemented using the scikit-learn 
\cite{pedregosa2011}. Figures were generated with Mayavi 
\cite{ramachandran2011} and nipy (\url{www.nipy.org}).

\section*{References}

\bibliography{restingstate}
\bibliographystyle{elsarticle-num-names}

\end{document}